\documentclass[reprint,noshowpacs,nofootinbib,fleqn]{revtex4-1}
\usepackage{graphicx,color}
\usepackage{amsmath}
\def\sl#1{#1\!\!\!\slash}
\def\smc#1{{\textsc{#1}}}

\begin{document} 
\title{\color{blue}\Large Distinctive Heavy Higgs Decays} 
\author{B.~Holdom}
\author{M.~Ratzlaff}
\affiliation{Department of Physics, University of Toronto, Toronto, Ontario, Canada  M5S1A7}
\begin{abstract}
For a heavy Higgs boson above the $t\bar t$ threshold we consider three decay modes that could compete with decay to $t\bar t$, even in the alignment limit of a 2-Higgs doublet model. The decays $H\to AZ$, $ Z'Z'$ and $\tau'\tau'$ show that heavy Higgs decays can be an effective probe of new physics both within and beyond this model.
\end{abstract}
\maketitle 
\squeezetable
A heavy Higgs $H$ may quite likely decay predominantly into $t\bar t$ when its mass is above this threshold, and this decay faces a daunting $t\bar t$ background. If the properties of the 125 GeV Higgs boson $h$ continue to converge to standard model expectations then it is less likely that some other decay modes of $H$ will be able to compete with the $t\bar t$ decay mode. This is because the suppression of these other decay modes is correlated with the standard-model-like properties of $h$ in the alignment limit of a 2-Higgs doublet model. The suppressed decay modes are $H\to WW$, $ZZ$, $hh$ and $A\to Zh$ where $H$ and $A$ are CP even and odd.

In this study we consider three decays that are not suppressed in the alignment limit.
\begin{itemize}
\item $H\to AZ\to b\bar b \ell^+\ell^-$
\item $H\to Z'Z'\to b\bar b \tau^+\tau^-$
\item $H\to \tau'\tau'\to \text{multi-}\ell+X$
\end{itemize}
The first is a standard decay within the 2-Higgs doublet model, while the other two show how heavy Higgs decay may serve as a discovery channel for new particles. The $Z'$ we consider does not couple to the first two families and so other production modes are suppressed. The $\tau'$ is a new heavy lepton with weak decays and is otherwise only weakly produced.

We assume the intermediate particles are on shell. For $H\to AZ$ this requires a suitably large mass splitting between $H$ and $A$. This splitting is determined by a certain combination of quartic couplings that survives in the alignment limit. In addition this splitting survives a $SU(2)_L\times SU(2)_R$ symmetry limit of the Higgs potential, unlike the mass splitting between $H$ and $H^\pm$. As described in \cite{Holdom:2014bla} there is some preference for the relations $m_A<m_H\approx m_{H^\pm}$.

As yet there has been no dedicated searches for any of these decay modes. In the first two cases we shall consider limits that can be set from the 13 TeV Run 2 at the LHC, and we describe cut based analyses that quite significantly increase these signals relative to background. The third case gives rise to a variety of multi-lepton final states such that strong limits can already be set with searches conducted in Run 1.

For the first two searches it will be important to realistically model $b$-tagging and $\tau$-tagging efficiencies in the detector simulation. We shall use \smc{delphes} 3 \cite{deFavereau:2013fsa} in two different ways for comparison purposes. \smc{checkmate} 1.1.13 \cite{Drees:2013wra}\footnote{A $\tau$-tagging bug was fixed in version 1.1.13.} uses a modified version of \smc{delphes} 3.0.10 that incorporates a set of efficiencies and other settings to more realistically model the \smc{atlas} detector. The $b$-tagging model is specified by a $b$-tag efficiency that we take to be 70\%. For $\tau$-tagging we choose the `medium' efficiency model along with a unit charge requirement. We then carry out our cut-based analysis within the \smc{checkmate} framework. We note that \smc{checkmate} has been validated against quite a large and growing number of analyses, many of which involve $b$-tagging and at least one where $\tau$-tagging plays an essential role. Thus the results we obtain from it can be considered fairly realistic.

Our second use of \smc{delphes} is based on stand-alone \smc{delphes} 3.1.2. We use the latest default \smc{atlas} detector card which includes a modeling of the jet energy scale. We also incorporate the efficiency for $b$-tag, $c$-jets and light jets that was used in a \smc{madanalysis} 5 \cite{Conte:2012fm,Conte:2014zja} implementation and validation \cite{Chalons} of a certain \smc{atlas} analysis. On the other hand $\tau$-tagging is more crudely implemented by stand-alone \smc{delphes} and we adopt the default 40\% constant efficiency. Then on the resulting \smc{delphes} root file we use \smc{madanalysis} 5 to apply the same cut based analysis as before.

\section*{\large $\color{blue}H\to AZ$}
We choose $M_H=500$ GeV and $M_A=300$ GeV as our benchmark point. These masses are larger than an analysis of $H\to AZ$ in \cite{Coleppa:2014hxa} , and we choose to work in the alignment limit of a 2HDM. The couplings of $H$ and $A$ to $t$ and $b$ are model dependent and could be larger or smaller than the corresponding $h$ couplings. For our benchmark point we set these $H$ and $A$ couplings to be the same as those for $h$. In this case, since the $HAZ$ coupling is well determined, we have $BR(H\to AZ)/BR(H\to t\bar t)\approx 0.7$ and $BR(A\to b\bar b)/BR(A\to gg)\approx 0.9$. Other competing decay modes can be small in the alignment limit and by ignoring them we obtain the values of $BR(H\to AZ)$ and $BR(A\to b\bar b)$ used in the following. We note that if $M_A$ is decreased then both of these branching ratios increase.

We use the \smc{feynrules} \cite{Alloul:2013bka} 2HDM model in \cite{FR1} (in the alignment limit) interfaced with \smc{herwig}++ 2.7.1  \cite{Bahr:2008pv} to generate the showered and decayed signal events. Heavy Higgs production proceeds through $gg\to H$ and $gg\to Hg$. Our event selection is a follows. Jets, including $b$-tags and $\tau$-tags, are required to have $p_T>20$ GeV and $|\eta|<2.5$. Muons (electrons) are required to have $p_T>20$ GeV and $|\eta|<2.4\;(2.47)$. For lepton isolation we require the $p_T$ sum of tracks within a cone of $0.3$ to be less than 0.16 of the lepton $p_T$. We also implement some standard overlap removals involving lepton pairs or lepton-jet pairs. We then apply the following sequence of cuts.
\begin{enumerate}
\item 2 $b$-jets
\item 2 $\ell$'s such that $M(e^+e^-)$ or $M(\mu^+\mu^-)$ is within $M_Z\pm 10$ GeV
\item $M(b\bar b\ell^+\ell^-)+M(b\bar b)>600$ GeV
\item $M(b\bar b\ell^+\ell^-)-M(b\bar b)>160$ GeV
\item $\Delta R(\ell^+\ell^-)<2$
\end{enumerate}

The dominant backgrounds are from $t\bar t$ and $b\bar b Z$ production. We use \smc{madgraph}5\_a\smc{mc}@\smc{nlo} \cite{Alwall:2014hca} to generate events for these processes at NLO which are then passed through \smc{herwig}++ for showering and decay. The \smc{mc}@\smc{nlo} method produces events with negative weight and we account for this in the analysis.

The cut flows in Table \ref{mine1} show that the cuts effectively suppress the $t\bar t$ background. For the production cross sections we take $\sigma(H)=7.6$ pb, $\sigma(t\bar t)=685$ pb and $\sigma(b\bar b Z)=87$ pb. The first is twice the LO value for $H+Hg$ production (to model higher order contributions) while the latter two are the NLO values. After incorporating the branching ratios as well we arrive at the numbers of the last row. For 100 fb$^{-1}$ the signal numbers are large, as is $S/\sqrt{B}$, so the significance is limited by systematics in the background estimate. Assuming that the standard model backgrounds can be determined through the use of control regions to within [15\%, 10\%], this would correspond to a signal significance of [1.7, 2.6] for \smc{checkmate} and [1.3, 2.0] for \smc{delphes}. These results may be scaled by factors that relate different choices of the production cross section and $BR(H\to AZ)$ and $BR(A\to b\bar b)$ to our choices. The results can be expected to improve if $M_A$ is smaller relative to $M_H$.
\linespread{1}\begin{table}[h]\centering\begin{tabular}{c|c|c|c|c|c|c|}\cline{2-7}  &\multicolumn{2}{|c|}{$H\to AZ\to b\bar b \ell\ell$}&\multicolumn{2}{|c|}{$t\bar t\to b\bar b\ell\ell\sl{E}_T$}&\multicolumn{2}{|c|}{$b\bar b Z\to b\bar b\ell\ell$}\\\cline{2-7}
 & CM & DE & CM & DE & CM & DE\\\cline{2-7}
 & $\;100000\;$ & 100000 & 100000  & 100000 & 100000 & 100000\\\hline 
\multicolumn{1}{ |c|  }{2 $b$-jets} &35166  & 39017 & 25049  & 25858 & 17757 & 16208\\\hline 
\multicolumn{1}{ |c|  }{$M_{\ell\ell}=M_Z\pm 10$ GeV} &17395  & 20077 & 795 & 874 & 7685 & 7203 \\\hline 
\multicolumn{1}{ |c|  }{$M_{bb\ell\ell}+M_{bb}>600$ GeV} & 11955 & 12407 & 121 & 109 &  1202 & 1426 \\\hline 
\multicolumn{1}{ |c|  }{$M_{bb\ell\ell}-M_{bb}>160$ GeV} & 11754 & 12245 & 32 & 33 & 800 & 987 \\\hline 
\multicolumn{1}{ |c|  }{$\Delta R_{\ell\ell}<2$} &10168  & 10539 & 15 & 17 & 584 & 808\\\hline\hline
\multicolumn{1}{ |c|  }{Events for 100 fb$^{-1}$} &1010  & 1047 & 480 & 543& 3404 & 4710\\\hline\end{tabular}
\caption{Cut flow for signal and two backgrounds at $\sqrt{s}=13$ TeV, comparing \smc{checkmate} and \smc{delphes}. $\ell=e$ or $\mu$.}
\label{mine1}\end{table}

\section*{\large $\color{blue}H\to Z'Z'$}
This decay can result if the standard heavy Higgs $H$ mixes with another scalar that is responsible for the mass of a $Z'$. We assume that the $Z'$ is hidden with respect to the first two families, so that up to the fermion mass mixings it couples only to the third family. Then the relative size of its coupling to $b$ and $\tau$ is important. For example if the $Z'$ coupling to $\tau$ is 3 times its coupling to $b$ then $BR(Z'\to\tau\tau)/BR(Z'\to b\bar b)=3$. In this case it would be advantageous to focus on the $4\tau$ final state and constrain it by simple multi-lepton searches. But a $Z'$ coupling only to the third family may arise if it is light remnant of some badly broken flavor interaction acting between families. Then it is more likely to have equal coupling to $b$ and $\tau$,\footnote{Anomalies can be canceled for example if the $Z'$ also coupled to a fourth family.} in which case $BR(Z'\to\tau\tau)/BR(Z'\to b\bar b)=1/3$. This makes the $b\bar b\tau\tau$ final state important to consider and this is our focus here.

Our goal is to see how well $BR(H\to Z'Z')$ can be constrained. We choose $M_H=500$ GeV and $M_{Z'}=170$ GeV as our benchmark point. For increasing $M_{Z'}$, $BR(H\to Z'Z')$ will decrease relative to $BR(H\to t\bar t)$ due to decreasing phase space.

We use a modification of the \smc{feynrules} model in \cite{FR2} interfaced with \smc{herwig}++ to generate the showered and decayed events. Our event selection is based on the same definitions of jets and isolated leptons as before. To reconstruct the $Z'$ mass from the two $\tau$'s we utilize the $\tau$-tagged jets as well as the leptons from the leptonic $\tau$ decays. The $\tau$'s are sufficiently boosted to assume that the visible and invisible decay products are collinear, and this allows the fractions $x_i$ of the two momenta that is visible to be determined. If this determination yields $0<x_i<1$ then we say that $\tau\tau$ is successfully reconstructed.

We apply the following sequence of cuts, with units in GeV.
\begin{enumerate}
\item at least 2 leptons and/or $\tau$-tags (leading two are used)
\item at least 2 $b$-jets (leading two are used)
\item successful $\tau\tau$ reconstruction
\item $M(b\bar b\tau\tau)  > M(b\bar b) + M(\tau\tau)-40$
\item $|M(\tau\tau)-M(b\bar b)-40| < 60$
\item $M(\tau\tau) > 130\text{ \& }M(b\bar b) > 90$
\item $-(p_{\tau_1}+p_{\tau_2}-p_{b_1}-p_{b_2})^2>200^2$
\end{enumerate}
Cuts 5 and 6 account for a downward shift in the observed $M(b\bar b)$ due to $\sl{E}_T$ from $b$ decays.

We generate the dominant $t\bar t$ background as before. Other backgrounds where $\tau\tau$ comes from a $Z$ are suppressed by the $M(\tau\tau) > 130$ GeV cut. Table \ref{mine2} shows our results, where the numbers in the last row use $BR(H\to Z'Z')=1$ and $BR(Z'Z'\to b\bar b\tau\tau)=3/8$ and the same production cross sections as before. If we assume a [15\%, 10\%] error in the determination of the background, then $BR(H\to Z'Z')=0.3$ for example would yield a signal significance close to [1.5, 2.2] from both \smc{checkmate} and \smc{delphes}. These results can again be scaled to account for different production cross sections or branching ratios.
\linespread{1}\begin{table}[h]\centering
\begin{tabular}{c|c|c|c|c|}\cline{2-5}  &\multicolumn{2}{|c|}{$H\to Z'Z'\to b\bar b \tau\tau$}&\multicolumn{2}{|c|}{$t\bar t\to b\bar b\ell\ell\sl{E}_T$}\\\cline{2-5}
 & CM & DE & CM & DE\\\cline{2-5}
 & $\;\;100000\;\;$ & 100000 & 100000  & 100000\\\hline 
\multicolumn{1}{ |c|  }{2 leptons/taus} &10406  & 8319 & 28180  & 28659\\\hline 
\multicolumn{1}{ |c|  }{2 $b$-jets} &2469  & 1824 & 7160 & 6892\\\hline 
\multicolumn{1}{ |c|  }{$\tau\tau$ reconstruction} & 1492 & 1238 & 1861 &  1783\\\hline 
\multicolumn{1}{ |c|  }{$M_{bb\tau\tau}>M_{bb}+M_{\tau\tau}-40$} & 1258 & 1070 & 499 & 513\\\hline 
\multicolumn{1}{ |c|  }{$|M_{\tau\tau}-M_{bb}-40|<60$} & 969 & 840 & 201 & 210\\\hline 
\multicolumn{1}{ |c|  }{$M_{\tau\tau}>130\text{ \& }M_{bb}>90$} & 880 & 738 &  91 & 102\\\hline 
\multicolumn{1}{ |c|  }{$-(p_{\tau_1}+p_{\tau_2}-p_{b_1}-p_{b_2})^2>200^2$} &760  & 622 & 39 & 33 \\\hline\hline
\multicolumn{1}{ |c|  }{Events for 100 fb$^{-1}$} &2166 & 1772 & 2804 & 2373 \\\hline\end{tabular}
\caption{Cut flow for signal and background at $\sqrt{s}=13$ TeV, comparing \smc{checkmate} and \smc{delphes}. The signal event numbers for 100 fb$^{-1}$ use $BR(H\to Z'Z')=1$. In $t\bar t\to b\bar b\ell\ell\sl{E}_T$ we let $\ell$ denote $e$, $\mu$, or $\tau$.}
\label{mine2}\end{table}

\section*{\large $\color{blue}H\to \tau'\tau'$}
Here we discuss how a heavy Higgs can provide strong constraints on new heavy leptons to which it couples. An example of a heavy Higgs with this decay was discussed in the context of a sequential fourth family \cite{Holdom:2014bla}. We thus focus on a new doublet of leptons $\nu', \tau'$ that have standard weak decay modes.  The $H\to \tau'\tau'$ decay may compete with the $t\bar t$ decay because the $\tau'$ mass may be of the same order as the top mass, and as shown in \cite{Holdom:2014bla}, the actual coupling ratio may deviate significantly from the mass ratio. We shall ignore the possible $H\to \nu'\nu'$ decay in this study.

In \cite{Holdom:2014rsa} we placed limits on the allowed  $\nu'$ and $\tau'$ mass combinations from production via $W$,$Z$,$\gamma$ only. Here we consider Higgs production as well for mass combinations that are not already excluded there. As in \cite{Holdom:2014rsa} our further limits will be obtained by recasting certain analyses based on 8 TeV data.

The final state from the heavy lepton decays will be strongly affected by the CKM mixing between the third and fourth family leptons and by the relative size of the $\tau'$ and $\nu'$ masses.  The mixing could be quite large ($\sim 0.1$) or it could be very small. We shall consider two cases which effectively maximize or minimize the possible numbers of light leptons in the final state.

In case A we consider very small mixing and $m_{\tau'}>m_{\nu'}+40$ GeV. This results in $\tau'\to \nu' W$ (not necessarily on-shell) and $\nu'\to\tau W$ and thus the following processes.
\begin{align*}
&H\to  \tau' \tau'\to \nu' W \nu' W\to WWWW\tau\tau \\
&Z\text{,}\gamma\to  \tau' \tau'\to WWWW\tau\tau\\
&W\to  \tau' \nu'\to WWW\tau\tau\\
&Z\to  \nu' \nu'\to WW\tau\tau
\end{align*}
Here we shall find that the four or more lepton final states provide quite stringent constraints on the $H$ production cross section times heavy lepton branching ratio. 

In case B we consider a mixing of 0.1 and $m_{\tau'}<m_{\nu'}+ 50$ GeV. Here the branching fraction for $\tau'\to \nu' W$ is no more than $10\%$. Then the dominant decays are $\tau'\to \nu_{\tau} W$ and $\nu'\to\tau W$, giving the following processes.
\begin{align*}
&H\to  \tau' \tau' \to W W\nu_{\tau} \bar{\nu}_{\tau} \\
&Z\text{,}\gamma\to  \tau' \tau'\to WW\nu_\tau\bar{\nu}_\tau\\
&W\to  \tau' \nu'\to WW\tau\nu_\tau\\
&Z\to  \nu' \nu'\to WW\tau\tau
\end{align*}
$\nu'\to \tau'W$ and $\tau'\to \nu' W$ are also included when they contribute. For case B we shall focus on the two lepton final states. The constraints will clearly be much weaker than case A and in fact these processes could have cross sections in the 2-4 pb range.

We use a \smc{feynrules} model interfaced with \smc{herwig}++ 2.7 to generate the showered and decayed events. We use \smc{delphes} and \smc{madanalysis} 5 to implement (recast) the experimental analyses of interest (in case A we follow the method described in \cite{Holdom:2014rsa}).  We set limits on the product $\alpha B$ where $B$ is the branching ratio for $H\to\tau'\tau'$ and $\alpha$ is the size of the $gg\to H$ cross section \emph{relative} to the one predicted for an $H$ with a SM-Higgs coupling to $t\bar t$. These latter cross sections are obtained from \cite{Heinemeyer:2013tqa}.

\subsection*{\color{blue} Case A}
\begin{figure}[t]
\centering
\includegraphics[scale=0.21]{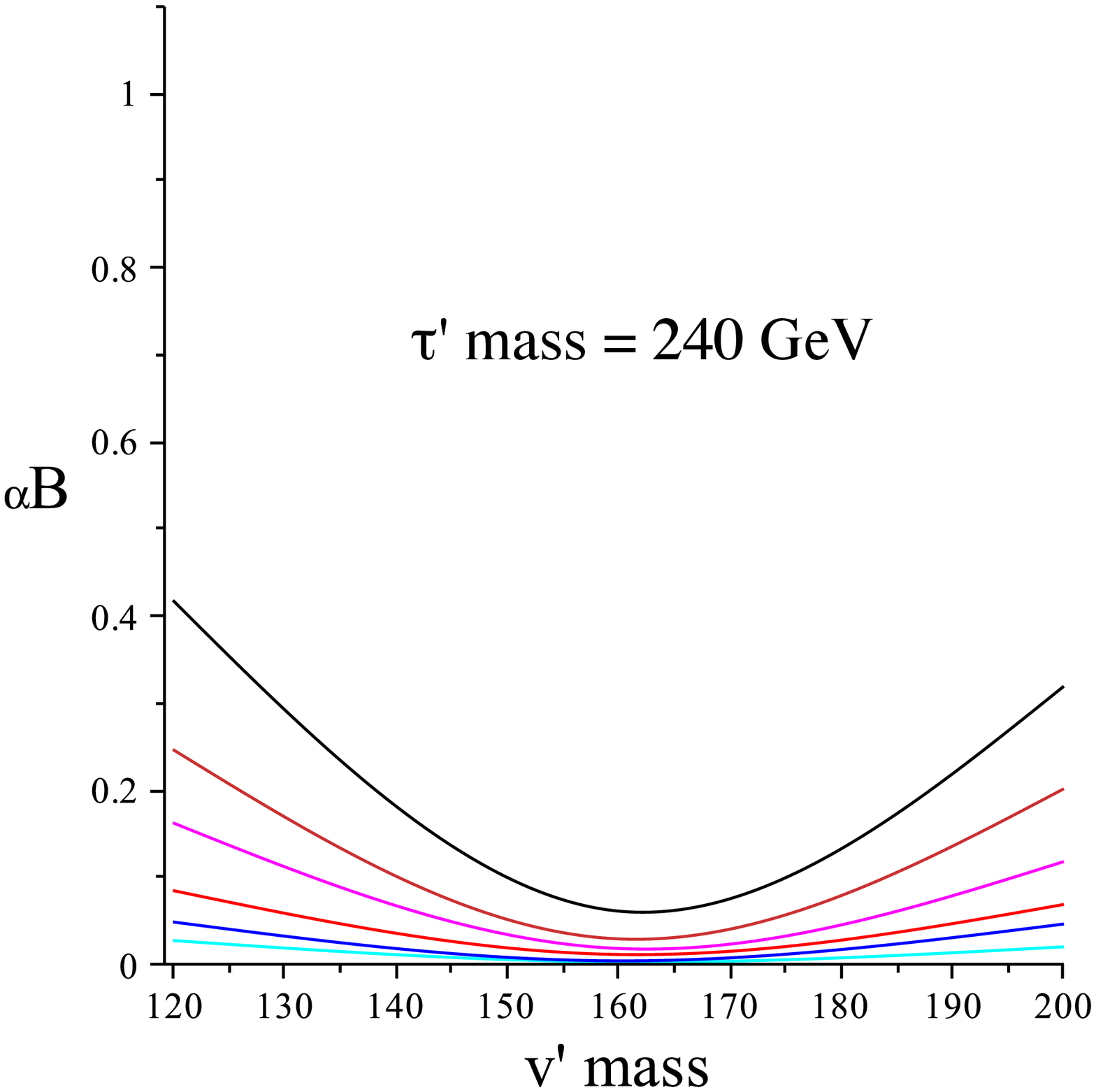}
\includegraphics[scale=0.21]{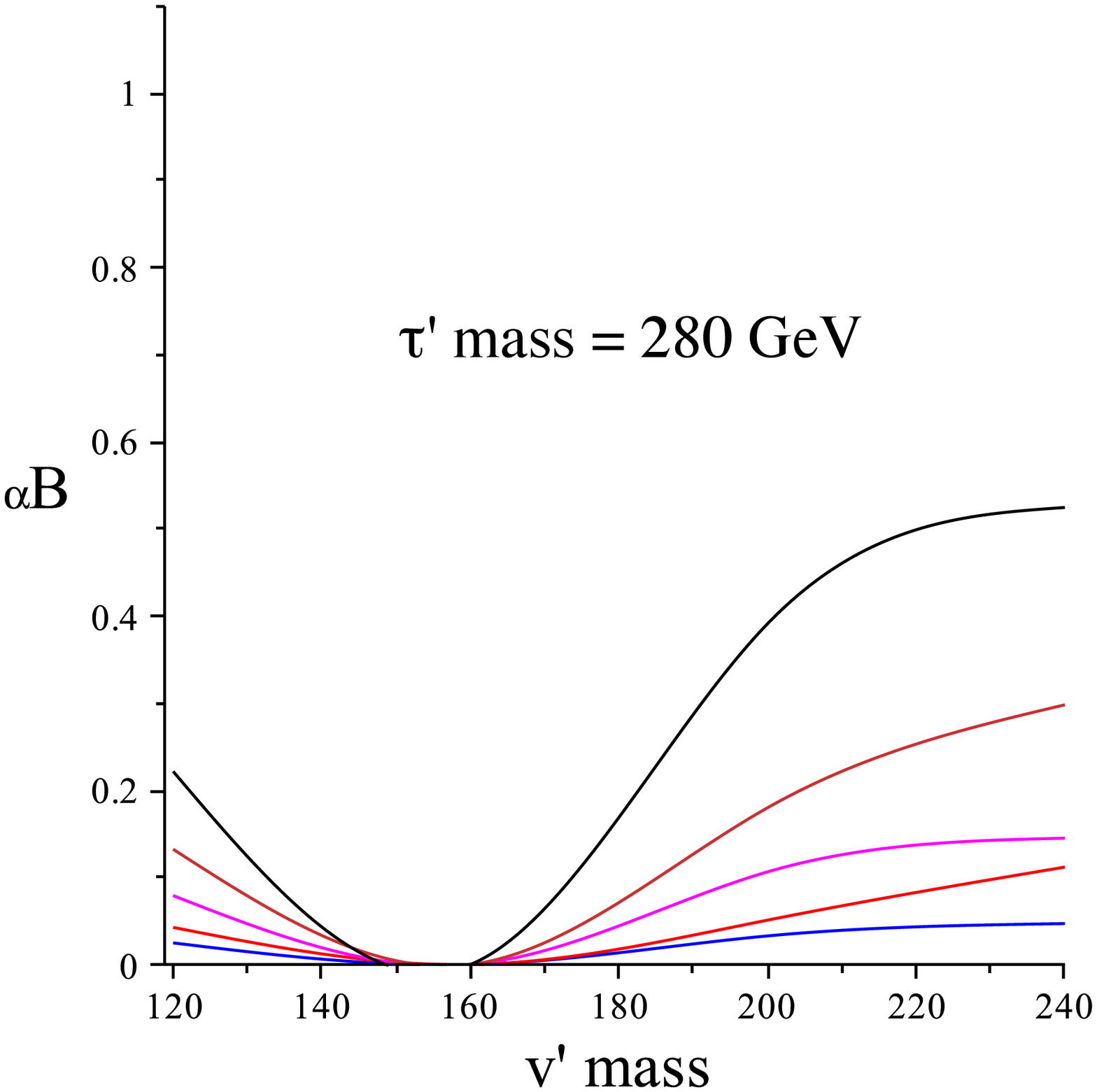}
\includegraphics[scale=0.21]{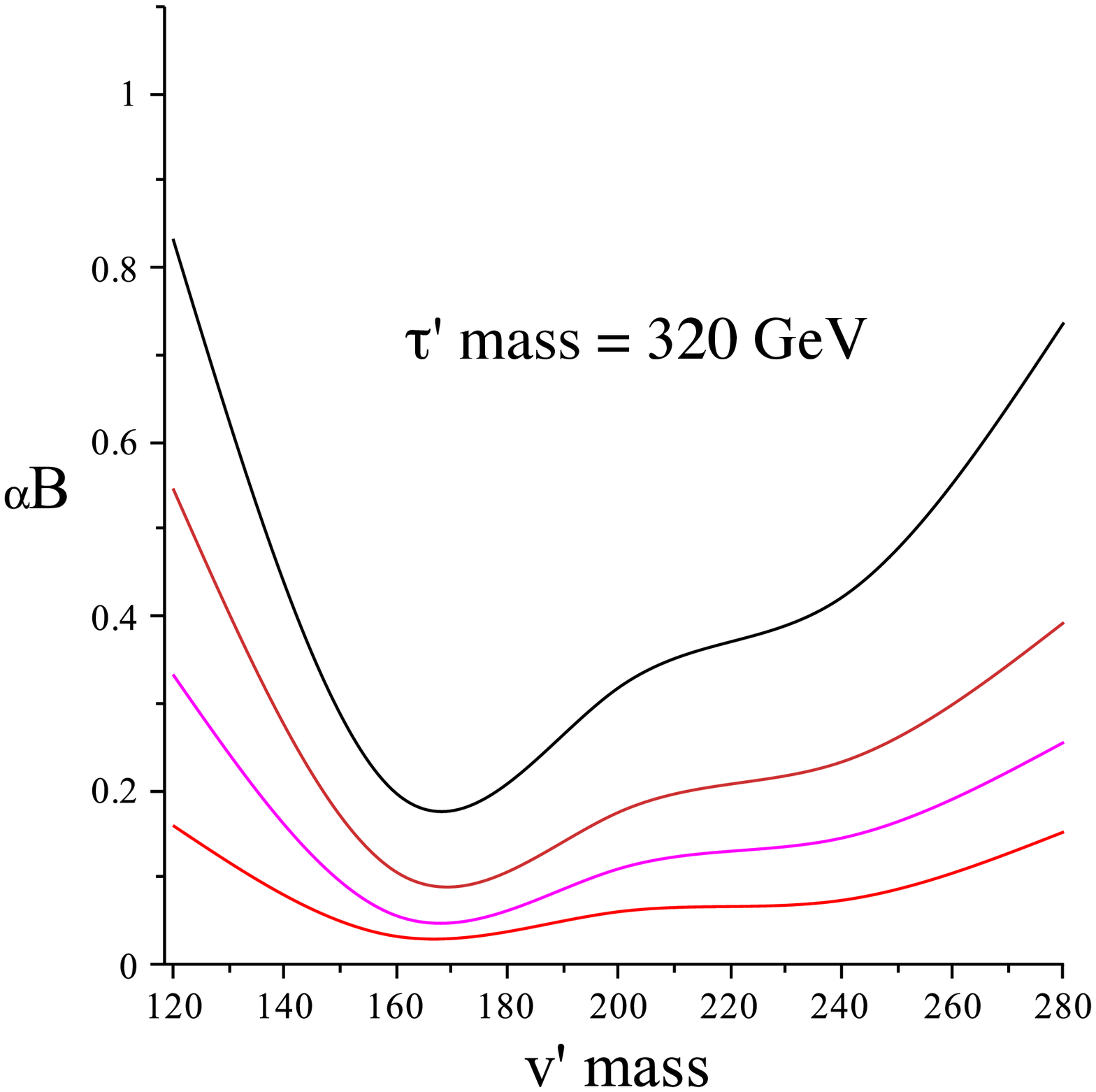}
\includegraphics[scale=0.21]{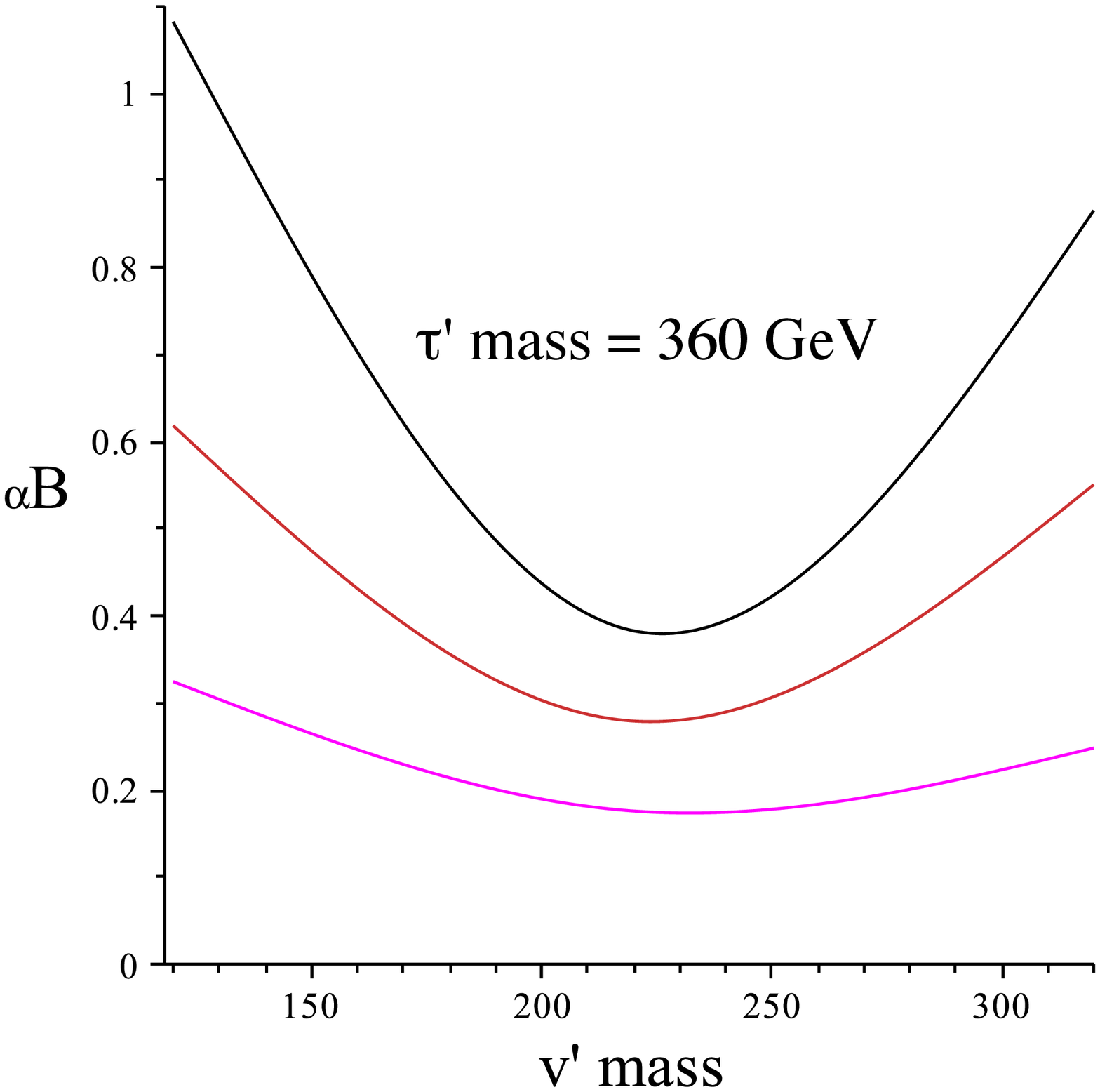}
\caption{Bounds on $\alpha B$ as a function of the heavy lepton masses. From high to low the curves are for $M_H=1000$ GeV (black), $M_H=900$ GeV (orange),  $M_H=800$ GeV (magenta), $M_H=700$ GeV (red), $M_H=600$ GeV (blue), $M_H=500$ GeV (cyan).}
\label{fit}
\end{figure}

We find that the most sensitive search is the \smc{atlas} search for four or more leptons and missing energy with $20.7$ fb$^{-1}$ \cite{ATLAS:2013qla}. We utilize all three ``off $Z$'' signal regions to set limits, since the efficiency times acceptance varies depending on the $H$ and heavy lepton masses. We include all the processes that we have listed for case A. The resulting 95\% CL constraints on $\alpha B$ are shown in Figure \ref{fit}.  The $\tau'$  and $\nu'$ mass combinations we consider are not excluded by \cite{Holdom:2014rsa} with the exception of the point $m_{\nu'}=160$ GeV and $m_{\tau'}=280$ GeV; for this point we set $\alpha B=0$.

These limits on $\alpha B$ are strong even for very massive $H$. The search has small backgrounds with no excess observed above the SM. The limits on the total number of allowed events in the three channels are $4.7$ in SR0noZa, $3.7$ in SR0noZb, and $7.5$  in SR1noZ. For the mass combinations we consider the number of events from production via $W$,$Z$,$\gamma$ in these three regions are 1.3-4.6, 1.2-3.6, and 2.9-7.2 respectively. For example an $800$ GeV Higgs with standard couplings has a production cross section of 110 fb while the total $W$,$Z$,$\gamma$ production cross section for $m_ {\nu'}=240$ GeV and $m_ {\tau'}=320$ GeV is $42$ fb.  The $H$ initiated process also has an efficiency times acceptance times branching ratio to four leptons roughly twice that of the $W$,$Z$,$\gamma$ process. In this example the strongest limit is in the SR1noZ channel where $W$,$Z$,$\gamma$ production results in $5$ events; after accounting for error in event generation we obtain $\alpha B<0.14$.

\subsection*{\color{blue} Case B}
The $\ell^+\ell^-+\sl{E}_T$ final state mimics the purely leptonic decay of $WW$, and so we may expect that the best constraints come from searches that target the $WW$ final state. We thus first consider the \smc{atlas} search for heavy Higgs with decay $H\to WW\to \ell\nu\ell\nu$ \cite{TheATLAScollaboration:2013zha} in the zero jet channel. The largest signals come from lepton masses not far above $100$ GeV, since then the production via $W$,$Z$,$\gamma$ also produces a significant number of events.

We verify our \smc{madanalysis} 5 cut flow by generating the standard model $WW$ background with \smc{herwig}++ and checking it against the \smc{atlas} cut flow. We find good agreement. The number of events produced via the processes we have listed for case B with $\alpha B=1$ are shown in Table \ref{mine3}. These numbers can be compared with the 95\% CL upper limit from the \smc{atlas} analysis of 97 events. The resulting upper limit on $\alpha B$ is the smallest (and close to unity) for the smallest lepton masses.
\linespread{1}\begin{table}[h]\centering
\begin{tabular}{|c|c||c|c|c|c|c|}\hline
&&$Z$,$W$,$\gamma$& H & H  & H & H \\
$m_{\nu'}$&$m_{\tau'}$& & $m_H=300$ & $m_H=400$  & $m_H=500$ & $m_H=600$\\\hline\hline
81&102&49&45&46&25&9\\
90&100&27&45&46&25&9\\
90&140&8&58&57&19&8\\
102&102&39&45&46&25&9\\
180&185&7&&60&24&9\\
240&240&2-3&&&32&15\\
290&290&1-2&&&&10\\\hline
110&105&20&45&46&25&9\\
160&102&22&45&46&25&9\\
280&140&12&58&57&19&8\\\hline
\end{tabular}
\caption{Event numbers contributed by heavy lepton production (case B) in the \smc{atlas} heavy Higgs search $H\to WW\to \ell\nu\ell\nu$, to be compared with the 95\% CL upper limit of 97 events. The last three rows have $m_{\nu'}>m_{\tau'}$.}
\label{mine3}
\end{table}

Next we consider the extent to which the production of heavy leptons could affect the measurement of the $WW$ production cross section. We thus compare our signal with the \smc{cms} $8$ TeV measurement with $3.5 $ fb$^{-1}$ \cite{Chatrchyan:2013oev} and the  \smc{atlas} $8$ TeV measurement with $20.3 $ fb$^{-1}$ \cite{ATLASWW}. Both searches have reported excesses with respect to standard model expectations, but it now appears that this may be due to poor theoretical understanding of the effect of jet vetoes \cite{Meade:2014fca,Jaiswal:2014yba}. In this case rather than comparing to a 95\% CL signal limit, we shall simply compare our signal strength directly to the size of the excesses.

We implement the cuts in \smc{madanalysis} 5 with the following exception. \smc{atlas} considers two types of missing energy, track and calorimeter based respectively, and then implements a cut on the angle between the two. We simply apply the observed effect of this cut in the \smc{atlas} cut flow to our signal cut flow. We also check our signal results by modeling the standard model $WW$ production using  \smc{herwig}++ scaled to the expected cross section. The number of events we generate in each signal region agrees within error with the number of $\mu^\pm e^\mp$,   $\mu^+ \mu^-$ and $e^+e^-$ MC events reported by the collaborations. Both $WW$ searches are also implemented in the \smc{checkmate} distribution. We have good agreement with the \smc{cms} \smc{checkmate} analysis while the \smc{altas} \smc{checkmate} cut flow for $WW$ is somewhat smaller than ours; normalizing that result to the data would effectively somewhat boost our signal strength.

In Table \ref{mine4} we give our predicted number of signal events contributing in the \smc{atlas} and \smc{cms} analyses assuming $\alpha B=1$. These numbers can be compared to the excesses reported, 152 events for \smc{cms} and 851 events for \smc{atlas}. For most choices of lepton masses we find that increasing $\alpha B$ sufficiently to account for these excesses would give an increase in the corresponding number of Table \ref{mine3} that would not be compatible with the $H\to WW\to \ell\nu\ell\nu$ search. Thus the effect of heavy leptons on the $WW$ cross section determination is expected to be subdominant to the present theoretical uncertainties, if the excesses are to be interpreted in this way.
\begin{table}[h]\centering
\begin{tabular}{|c|c||cc|cc|cc|cc|cc|}\hline
&& \multicolumn{2}{|c|}{ $Z$,$W$,$\gamma$} & \multicolumn{2}{|c|}{ H} & \multicolumn{2}{|c|}{ H}  & \multicolumn{2}{|c|}{ H}& \multicolumn{2}{ |c|}{ H}\\
$$&$$& \multicolumn{2}{|c|}{ $$} &\multicolumn{2}{|c|}{ $m_H=300$} & \multicolumn{2}{ |c|}{ $m_H=400$}  & \multicolumn{2}{|c|}{ $m_H=500$}& \multicolumn{2}{ |c|}{ $m_H=600$}\\
$m_{\nu'}$&$m_{\tau'}$&\textsc{cms}&\textsc{atlas}&\textsc{cms}&\textsc{atlas}&\textsc{cms}&\textsc{atlas}&\textsc{cms}&\textsc{atlas}&\textsc{cms}&\textsc{atlas}\\\hline\hline
81&102&39&190&40&184&37&155&17&65&8&29\\
90&100&40&182&40&260&37&155&17&65&8&29\\
90&140&18&82&53&184&46&187&19&65&8&28\\
102&102&35&165&40&184&37&155&17&65&8&29\\
180&185&6&23&&&51&191&22&85&9&30\\
240&240&$\sim2$&$\sim8$&&&&&26&86&9&36\\
290&290&$\sim1$&$\sim5$&&&&&&&9&33\\
\hline
110&105&26&121&40&184&37&155&17&65&8&29\\
160&102&23&98&40&184&37&155&17&65&8&29\\
280&140&7&32&53&260&46&187&19&65&8&28\\
\hline
\end{tabular}
\caption{Event numbers contributed by heavy lepton production (case B) in the \smc{cms} and \smc{atlas} $WW$ cross sections analyses, to be compared with the reported excesses of 152 and 851 events respectively.}
\label{mine4}
\end{table}

We note that $H\to \tau' \tau'$ produces a larger high energy tail in the lepton $p_T$ distribution, at least for some mass combinations, and this could in the future provide a way of distinguishing the  signal from that of $WW$ production.

For both cases A and B we can expect updates from Run 2 of the searches that we have used from Run 1, and it will be straightforward to recast these new results to further constrain heavy Higgs and new leptons.

\begin{acknowledgments}
This research was supported in part by the Natural Sciences and Engineering Research Council of Canada.
\end{acknowledgments}
\linespread{1}
 
\end{document}